\documentstyle[prl,aps,epsfig]{revtex}
\bibstyle{unsrt}
\tighten
\begin{document}

\twocolumn[\hsize\textwidth\columnwidth\hsize\csname
@twocolumnfalse\endcsname
\draft

\title{The Quantum-Classical Transition in Nonlinear Dynamical Systems}

\author{Salman Habib$^1$, Kurt Jacobs$^1$, Hideo Mabuchi$^2$, Robert
Ryne$^3$, Kosuke Shizume$^4$, and Bala Sundaram$^5$}    
\address{$^1$ T-8, Theoretical Division, MS B285, Los Alamos National
Laboratory, Los Alamos, New Mexico 87545}  
\address{$^2$ Mail Code 12-33, California Institute of Technology,
Pasadena, CA 91125}
\address{$^3$ LANSCE-1, LANSCE Division, MS H817, Los Alamos National
Laboratory, Los Alamos, NM 87545}  
\address{$^4$University of Library and Information Science, 1-2
Kasuga, Tsukuba, Ibaraki 305, Japan} 
\address{$^5$Department of Mathematics \& Graduate Faculty in Physics,
 CSI-CUNY, Staten Island, NY 10314}  
\date{\today}
\maketitle

\begin{abstract}
Viewed as approximations to quantum mechanics, classical evolutions
can violate the positive-semidefiniteness of the density matrix. The
nature of this violation suggests a classification of dynamical
systems based on classical-quantum correspondence; we show that this
can be used to identify when environmental interaction (decoherence)
will be unsuccessful in inducing the quantum-classical transition. In
particular, the late-time Wigner function can become positive without
any corresponding approach to classical dynamics. In the light of
these results, we emphasize key issues relevant for experiments
studying the quantum-classical transition.
\end{abstract}

\pacs{{PACS Numbers: 03.65.Yz, 05.45.-a}  
\hfill   LAUR-00-4046}
\vskip1pc]

In recent years much effort has been expended, both theoretically and
experimentally, to explore the transition from quantum to classical
behavior in a controlled way. In this context, the interaction of
trapped, cold atoms with optical potentials, both time-dependent and
independent, has become a topic of considerable interest and activity
\cite{exp,latt}. The experimental ability to systematically study
dissipative quantum dynamics of nonlinear systems is an exciting new
area where the frontier between classical and quantum mechanics may be
carefully examined.

In this Letter we wish to explore some of the key qualitative features
of the quantum-classical transition. We establish that with $\hbar$
fixed at a finite value, and classical dynamical evolution equations
for phase space distribution functions viewed as approximations to the
underlying quantum equations, classical Liouville and Master equations
violate the quantum constraint of positive-semidefiniteness of the
density matrix: We refer to this property of the density matrix as
`rho-positivity'. There is thus a global obstruction to the classical
limit arising directly from quantum evolutions. We argue below that
rho-positivity violation can (1) serve as a guide in classifying
dynamical systems with regard to classical-quantum correspondence:
weak violation as Type I, strong violation as Type II, and (2) explain
robustness to decoherence in the sense of avoidance of the classical
limit as exemplified by dynamical localization in the (open-system)
quantum delta kicked rotor (QDKR). Our results impact directly on the
interpretation and design of experiments to test various aspects of
the quantum-classical transition.

As described in more detail below, certain nonclassical aspects of the
dynamics of the QDKR turn out to be stable to decohering effects of
external noise and decoherence due to spontaneous emission. A
dynamical description in terms of the Wigner function leads to two
alternatives to explain this stability: (1) diffusion in the quantum
Master equation is simply not efficient at suppressing quantum
interference terms present in the Wigner function, or (2) the much
more intriguing possibility that the diffusion is successful at
decohering the Wigner function, i.e., interference terms are
suppressed and the Wigner function is (almost) everywhere positive,
yet the late-time distribution is not the solution of the
corresponding classical Fokker-Planck equation. We find that the
second possibility is the one actually realized, and show how it
arises as a consequence of the fact that the classical Fokker-Planck
equation violates rho-positivity, while the quantum Master equation
does not.

The singular nature of the classical limit $\hbar\rightarrow 0$ in
quantum mechanics has been appreciated for a long time. However, what
has not been stressed sufficiently is the reason for this singular
behavior, that classical dynamics violates unitarity and
rho-positivity, and thus $\hbar=0$ cannot be connected smoothly to
$\hbar\rightarrow 0$. A simple example suffices to make this point
clear. Let us consider as initial condition a pure Gaussian state. Let
us suppose that we evolve the corresponding (positive) Wigner function
classically in some nonlinear potential (for linear systems classical
and quantum dynamics are identical \cite{ah}), then the distribution
becomes no longer Gaussian, but is still positive-definite. Three
possibilities now present themselves: the evolved object can be
interpreted as (1) a pure quantum state (unitarity is preserved), (2)
a mixed quantum state (rho-positivity is preserved), and (3) cannot be
interpreted as a quantum state (rho-positivity is violated). The first
possibility can be dismissed using Hudson's theorem: the {\em only}
pure state with positive Wigner function is a Gaussian state with a
(necessarily) Gaussian Wigner function \cite{rh}. But our distribution
is non-Gaussian. As to the second, we first note that the phase space
integral of any function of the phase space distribution is preserved
under a Liouville flow. In particular $\int f^2(x,p)dxdp$ remains
constant. For Wigner functions this quantity is proportional to
$Tr\rho^2$ which is a direct measure of whether a state is mixed or
not -- since this measure cannot change, the evolved object is not
interpretable as a mixed state. Thus we are forced to the third
alternative, that the evolved object cannot be interpreted as a
quantum state at all, i.e., the Weyl transform of the evolved
classical distribution yields a `classical density matrix' which is
non-rho-positive \cite{shcl}.

The above analysis makes it clear that the classical Liouville
equation can never arise as a formal limit of quantum theory. However,
all real experiments deal with open systems, i.e., systems interacting
with their environment, of which the particular case of a measuring
apparatus (necessary to deduce classical behavior) is an important
example. Quantum decoherence and conditioned evolution arising as a
consequence of such system-environment couplings and the act of
measurement and observation provide a natural pathway to the classical
limit as has been quantitatively demonstrated, {\em e.g.}, in
Ref. \cite{bhj}. Thus, it is important to inquire into the role of
rho-positivity violation in this context: When is it important, and
when not?

Conditions have been previously derived that apply to the extraction
of (noisy) classical trajectories via continuous quantum measurement
\cite{bhj}. Once these (strict) conditions are satisfied (typically in
the macroscopic limit $\hbar\ll S$, where $S$ is the system action),
measurement induces classical behavior, and in this regime all systems
are therefore Type I.  However, when these conditions are not
satisfied, as is the case in most current experiments, the differences
are indeed important.  The key point is that, under continuous
measurement, Type I systems can violate the classicality conditions in
the sense that individual classical trajectories cannot be extracted,
yet expectation values are close to the classical results, whereas in
Type II systems, violation of the classicality condition also implies
a violation of correspondence at the level of expectation values. We
will demonstrate this for the QDKR below.

A quantum Master equation representing a weakly coupled, high
temperature environment often utilized in studies of decoherence is
\begin{eqnarray}
{\partial \over \partial t}f_W&=&L_{cl}f_W+ L_q
f_W + D{\partial^2 \over \partial p^2}f_W;
\label{qme}\\
L_{cl}&\equiv&-{p\over m}{\partial\over \partial
x}+ {\partial V\over\partial x}{\partial \over \partial
p}, \label{lcl}\\ 
L_q &\equiv& \sum_{\lambda~odd}{1\over\lambda!}\left({\hbar\over
2i}\right)^{\lambda-1}{\partial^{\lambda}V(x)\over\partial
x^{\lambda}}{\partial^{\lambda}\over\partial p^{\lambda}}.
\label{lq}
\end{eqnarray}

An unraveling of the weakly-coupled, high temperature environment,
Master equation (\ref{qme}) is provided by a continuous measurement of
position. This process is described by a stochastic Master equation
for the density matrix $\rho (t)$, conditioned on the measurement
record $\langle X \rangle + \xi(t)$ with $\xi(t) \equiv (8\eta
k)^{-1/2}dW/dt$, \cite{measx}
\begin{eqnarray}
\rho(t+dt) &=& \rho - (\frac{i}{\hbar} [H,\rho] - k [X,[X,\rho]]) 
                  \mathbin{} dt 
               \nonumber\\
           && {} + \sqrt{2\eta k} \mathbin{} ( [X,\rho]_+ - 
                2 \rho \mathop{\rm Tr} \rho X ) \mathbin{} dW \,,
\label{sme}
\end{eqnarray}
where $k$ is a constant specifying the strength of the measurement,
$\eta$ is the measurement efficiency and is a number between 0 and 1,
and $dW$ is a Wiener process, satisfying $(dW)^2=dt$.  When $\eta=1$,
the evolution preserves the purity of the state and can be rewritten
in a way which allows it to be understood as a series of diffuse
projection measurements~\cite{cqm2}. Averages over the resulting
Schr\"odinger trajectories reproduce expectation values computed using
the reduced density matrix $\rho$ or the corresponding Wigner function
$f_W(x,p)$ obtained from solving the Master equation (\ref{qme}). The
strength of the measurement is related to the diffusion coefficient of
Eq. (\ref{qme}) by $D=\hbar^2 k$.

When the diffusion constant $D=0$, Eq. (\ref{qme}) is just the quantum
Liouville equation for the closed system. Note that the linearity of
the quantum Liouville equation implies that in order for the evolution
to be unitary, $L_q$ cannot be unitary since $L_{cl}$ is not (the sum
$L_{cl}+L_q$ is unitary but not the operators separately). The
familiar heuristic argument for obtaining the classical limit from the
quantum Master equation is that the diffusion term smooths out the
interference effects generated by $L_q$ in such a way that quantum
corrections to the classical dynamics are much reduced. It has also
been argued, that at finite $\hbar$, the appropriate limiting case of
the quantum Master equation is in fact the classical Fokker-Planck
equation [setting $L_q=0$ in Eqn. (\ref{qme})] rather than the
classical Liouville equation \cite{hsz}. In any case, one immediately
appreciates that if {\em either of the classical equations are
strongly rho-positivity-violating, i.e., are Type II} then this
implies the existence of compensatory `large' quantum corrections in
the quantum Master equation, and hence the above heuristic argument
must fail: $L_q$ is responsible for more than just the generation of
interference fringes in the Wigner evolution.

Previous work has already suggested the possibility that closed
dynamical systems may be roughly divided into two types depending on
the (dynamical) classical-quantum correspondence as follows: (1) Type
I systems in which quantum expectation values and classical averages
track each other relatively closely as a function of time
\cite{hsz,close}, {\em e.g.}, the driven Duffing oscillator with
Hamiltonian,
\begin{equation}
H_{\hbox{duff}}=p^2/2m + B x^4 - A x^2 + \Lambda x \cos(\omega t),         
\label{hduff}\\
\end{equation}
and (2) Type II systems in which the quantum and classical averages
diverge sharply after some finite time, e.g., dynamical localization
in the QDKR \cite{dloc}. The QDKR Hamiltonian is
\begin{equation}
H_{\hbox{dkr}}={1\over 2}p^2+\kappa \cos q \sum_n\delta(t-n).          
\label{hdkr}
\end{equation}
We solved the classical and quantum Master equations corresponding to
Eqs. (\ref{hduff}) and (\ref{hdkr}) using a high-resolution spectral
solver implemented on parallel supercomputers. The solver explicitly
respects rho-positivity conservation.

We verified that in both the QDKR and the Duffing oscillator numerical
examples discussed below the localization condition \cite{bhj}
necessary to obtain classical trajectories was violated. The relevant
condition is $8\eta k\gg (\partial^2_x F /F)\sqrt{\partial_x F/2m}$
where the force $F$ and its derivatives are evaluated at typical
points in phase space. For both cases we have in fact, $8\eta k\sim
(\partial^2_x F /F)\sqrt{\partial_x F/2m}$ thus localization does not
occur (direct numerical solution of the corresponding stochastic
Schr\"odinger equation confirms this result) and, as discussed
earlier, a meaningful distinction between Types I and II is
possible. As the value of $\hbar$ is reduced (with $D$ fixed and
non-zero) one does expect an approach to the classical limit
\cite{bhj}, though the trajectory in the space of $D$ and $\hbar$ need
not be simple \cite{bhjs}.

\begin{figure}
\centerline{\epsfig{file=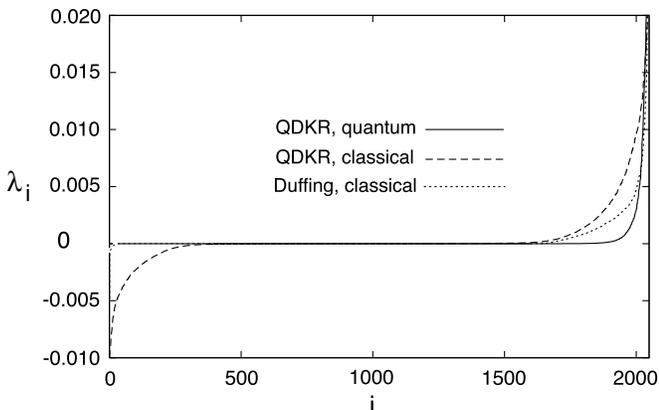,width=\hsize}}
\caption{Eigenvalues of the quantum density matrix (solid) and the
classical approximation (long-dashed) computed from the quantum and
classical Master equation evolutions for the QDKR at $t=6$. Also shown
(short-dashed) is the classical result for the Duffing oscillator at
$t=10$.}
\label{fig1}
\end{figure}

Our numerical code returns us the classical distribution function, the
quantum density matrix and the Wigner function as functions of
time. We then numerically solve for the eigenvalues of the quantum
density matrix and the eigenvalues of the Weyl transform of the
classical phase space distribution (the `classical density
matrix'). Results of one such computation are displayed in Fig. 1 for
the QDKR (Type II) and Duffing system (Type I). For the QDKR, initial
conditions are pure Gaussian Wigner functions characterized by the
position width $\Delta x=2.5$, momentum width $\Delta p=1$, centered
on the point $(x,p)=(0,0)$, and with $\hbar=5$ and $\kappa=10$. The
diffusion coefficient $D=0.1$, corresponding to $k=0.004$. The
horizontal axis refers to the index $i$ corresponding to the
eigenvalues $\lambda_i$, which are themselves plotted on the vertical
axis. The solid line is a result from a numerical solution of the
quantum Master equation, as expected all eigenvalues are positive (the
pure initial state has one eigenvalue equaling unity, the rest being
zero). The dashed line is the corresponding result from the classical
Fokker-Planck equation, which is characterized by a strong
contribution from negative eigenvalues. It is thus clear that the true
quantum density matrix and that provided by the classical
approximation are in fact quite different. In contrast, results from
classical Duffing calculations show a very small contribution from
negative eigenvalues [Parameter values in the particular case shown in
Fig. 1 were $m=1$, $A=10$, $B=0.5$, $\Lambda=10$, $\omega=6.07$,
$\Delta x=0.05$, $\Delta p=1$, $(x,p)=(-3,8)$, $\hbar=0.1$, $D=0.02$.]
These results show how rho-positivity violation may be used to
distinguish the two types of dynamical systems. An important point to
emphasize is that it is sufficient to only carry out the
classical dynamical calculation in order to classify a dynamical
system as being Type I or II. (The initial condition must of course be
a Wigner function.) Also, it should be clear that non-violation of
rho-positivity is a necessary but not sufficient condition for
quantum-classical correspondence in terms of agreement of expectation
values.

\begin{figure}
\centerline{\epsfig{file=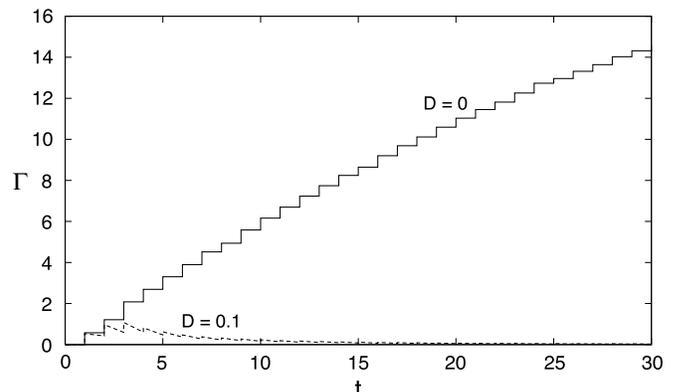,width=\hsize}}
\caption{The Wigner function negativity measure $\Gamma$ as a function
of time for $D=0$ and $D=0.1$ for the QDKR.}
\label{fig2}
\end{figure}

It is well-known that dynamical localization in the QDKR can be
destroyed (in the sense that $\langle p^2(t)\rangle$ no longer
saturates at late times) by coupling to external noise or to
dissipative channels (e.g., spontaneous emission)
\cite{locdiff}. However, what is important to note is that even in the
presence of quite strong coupling to these decohering channels, the
evolution does not go over to the classical one, and in this sense the
QDKR is quite different from the Duffing system investigated in
Ref. \cite{hsz}. In addition to numerically solving the Master
equation (\ref{qme}) we have investigated in detail the effects of
including amplitude and phase noise, timing jitter in the kicked
system, and carried out more realistic simulations taking into account
the effects of spontaneous emission. Stability to decoherence in the
sense above was manifest in all of these cases. Since we have
established that the QDKR is a Type II system (Fig. 1), this behavior
is essentially forced: as long as the classical evolution strongly
violates rho-positivity, it is impossible for the full evolution to
ever become close to the classical limit as the quantum corrections
must always be concomitantly large. The question remains whether the
resulting Wigner function at least has a classical interpretation. In
order to investigate this we computed as a function of time, the
quantity $\Gamma=\int dxdp(|f_W|-f_W)$, which provides a global
measure of negativity of the Wigner function. With $D=0$, one sees
that $\Gamma$ increases monotonically as the Wigner function develops
the expected oscillatory structure as a consequence of quantum
interference in phase space. When $D\neq 0$, diffusion in phase space
wipes out the interference and produces an essentially positive
distribution which one may interpret classically. However, because
rho-positivity must be maintained, classical evolution cannot
connect two such positive distributions. Thus, in Type II systems
decoherence can be successful in rendering the Wigner distribution
positive, but yet not lead to a classical limit. We note that in NMR
systems there is an interesting question regarding when classical
evolution of spins can reproduce quantum evolutions connecting spin
states that have no entanglement and thus may be interpreted
classically \cite{caves}. We have shown that a similar situation can
arise even in single-particle evolution where entanglement is not an
issue.

Recent experiments have attempted to directly address the issue of
environment-induced decoherence in the QDKR, in the context of cold
atom optics \cite{exp}. Despite some complications stemming from
non-ideal realizations, the results indicate that classical and
quantum evolutions agree only at inordinately large noise levels. Note
that in these experiments, parametric noise or spontaneous emission
was used as the decohering mechanism. (The non-selective Master
equation for atomic motion in far-detuned laser light has a similar
form to that of a particle subjected to continuous position
measurement. However, arguments can be made that only the weak
decoherence regime can be accessed in this manner.) The parameter
values in our numerical work are close to those actually used in the
experiments thus, as with our simulations, the experiments are not
carried out in a classical regime in the sense of Ref. \cite{bhj}. And
since we have demonstrated the strongly Type II nature of the QDKR, it
follows immediately that to observe true classical behavior, either
the current experiments have to switch to a Type I system or employ lower
values of $\hbar$. Simply increasing the measurement constant $k$, or
equivalently $D$, while it produces localization, adds noise into the
trajectory which must be kept small in order to achieve the classical
limit. This final condition requires a reduction in $\hbar$ as $k$ is
increased \cite{bhj}.


The authors acknowledge helpful discussions with Tanmoy Bhattacharya,
Doron Cohen, and Andrew Doherty. The work of BS was supported by the
National Science Foundation and a grant from the City University of
New York PSC-CUNY Research Award Program. Large-scale numerical
simulations were carried out at the Advanced Computing Laboratory
(ACL), LANL and at the National Energy Research Scientific Computing
Center (NERSC), LBNL.

\end{document}